\documentstyle[draft]{mn}
\input epsf
\input psfig.sty

\begin{document}
%\emergencystretch = 40pt
\newcommand{\mum}{$\,\mu$m}
\newcommand{\pho}{\phantom{1}}
\newcommand{\phm}{\phantom{-}}
\newcommand{\phe}{\phantom{^{\rm \,e}}}

\journal{Preprint UBC-COS-99-03, astro-ph/9909092}

\title[A search for the submillimetre counterparts to Lyman break galaxies]
{A search for the submillimetre counterparts to Lyman break galaxies}

\author[S.C.~Chapman et al.]
       {Scott C.~Chapman,$^{1,2}$ 
	Douglas  Scott,$^{1}$
	Charles C.~Steidel,$^{3}$
 	Colin Borys,$^{1}$
\newauthor Mark Halpern,$^{1}$
	Simon L.~Morris,$^{4}$
	Kurt L.~Adelberger,$^{3}$
	Mark Dickinson,$^{5}$
\newauthor
	Mauro Giavalisco$^{5}$ 
	and Max Pettini$^{6}$\\
	\vspace*{1mm}\\
        $^1$Department of Physics \& Astronomy,
	University of British Columbia,
	Vancouver, B.C.~V6T 1Z1,~~Canada\\
	$^2$Present address: Observatories of the Carnegie Institution of
	Washington, Pasadena, CA 91101,~~U.S.A.\\
	$^3$Palomar Observatory, Caltech 105-24,
	Pasadena, CA 91125,~~U.S.A.\\
	$^4$National Research Council of Canada,
	Herzberg Institute of Astrophysics,
	Victoria, B.C.~V8X 4M6,~~Canada\\
	$^5$Space Telescope Science Institute,
	3700 San Martin Drive, Baltimore, MD 21218,~~U.S.A.\\
	$^6$Institute of Astronomy, Madingley Road,
	Cambridge CB3 0HA,~~U.K.
        }

%\date{\fbox{\sc Submitted version}}
%\date{Accepted ... ; Received ... ; in original form ...}
\date{Submitted to Mon.~Not.~R.~astr.~Soc.}
\pubyear{1999}
\pagerange{000--000}
\maketitle

\begin{abstract}
We have carried out targetted sub-mm observations as
part of a programme to
explore the connection between the rest-frame UV and far-IR
properties of star-forming galaxies at high redshift, which is
currently poorly understood.  On the one hand the Lyman break
technique is very effective at selecting $z\,{\sim}\,3$ galaxies.  On the
other hand `blank field' imaging in the sub-mm seems to turn
up sources  routinely, amongst which some are star forming galaxies at similar
redshifts.  Already much work has been done searching for optical
identifications of objects detected using the SCUBA instrument. Here
we have taken the opposite approach, performing sub-mm photometry for a
sample of Lyman break galaxies 
whose UV properties imply high star formation rates.

The total signal from our Lyman break sample is undetected in the sub-mm, 
at an rms level of ${\sim}\,0.5\,$mJy, which
implies that the population of Lyman break galaxies does not constitute
a large part of the
recently detected blank-field sub-mm sources.  However, our one detection
suggests that with reasonable SCUBA integrations we might expect to
detect those few LBGs that are far-IR brightest.
\end{abstract}


\section{Introduction}

Using a variety of techniques in different wavebands, the detailed
study of young galaxies is being pushed to higher redshifts.  The
method of selecting high redshift galaxies using multi-colour
broadband observations of the rest-frame UV stellar
continuum has been successfully applied to ground-based surveys, and
also to the Hubble Deep Field (HDF).  In particular, the absence of
emission in the $U$-band at $z\,{\sim}\,$3 due to the
presence of the Lyman break feature has been very effective
\cite{SteHam,StePH,Ste98}, and is usually referred to as the Lyman
break technique (e.g.~Steidel et al.~1996a).  A full characterization
of the properties of the population of galaxies chosen in this way,
called Lyman break galaxies (LBGs), will likely provide answers to
some key questions of galaxy formation and evolution, particularly
those dealing with the murky role of dust at high redshift.

The colours of Lyman break galaxies are observed to be redder than
expected for dust-free star-forming objects.  The spectral slope of
the ultraviolet continuum and the strength of the H$\beta$ emission
line suggest that some interstellar dust is already present in these
young galaxies, and that it attenuates their UV luminosities by a
factor of ${\sim}\,~4.5$, although factors of as much as $100$ are implied
for some LBGs (Steidel et al.~1998).  In the most extreme objects this
implies star formation rates (SFRs) of several times
$100\,{\rm M}_\odot\,{\rm yr}^{-1}$, reaching as high as
$1000\,{\rm M}_\odot\,{\rm yr}^{-1}$ in some cases.
The resulting revisions to the global
star formation rate contributed by galaxies at redshifts $z\,{>}\,2$
can be significant.  However, the prescription for a `correct'
de-reddening is still unknown at present (see for example Meurer et
al.~1997).  This leads to an uncertainty in the estimates of dust
obscuration in LBGs derived from rest frame ultraviolet spectra (see
e.g.~Pettini et al.~1998; Calzetti et al.~1996).

The role dust plays in these young galaxies is not clear cut, since
redder spectra can also occur from an aging population, or from an
initial mass function (IMF) deficient in massive stars (see
e.g.~Bruzual \& Charlot~1993).  Recent studies of nearby star-forming
galaxies (Tresse and Maddox 1998) have concluded that the extinction
at a wavelength of $2000\,$\AA\ is typically 1.2 mag.  The situation at
earlier epochs is unclear, but
there is no need to assume that at higher redshifts
an increasing fraction of the star formation activity takes
place in highly obscured galaxies.
%(although high-$z$ galaxies could be quite different).
The precise amounts of reddening
required, and its interpretation, are therefore still open questions.
 
Observations using a sensitive new bolometric array, SCUBA on the JCMT
(described in section~\ref{sec:observations}), have discovered a
population of submillimetre detected galaxies
\cite{SIB,Hughes,Barger,Eetal,Holl98,BKIS} which might push up the global
SFR at $z\,{\sim}\,3$ by a factor of perhaps five (see for example
Smail et al.~1997).  Identifying the optical counterparts to these
galaxies, however, is a non-trivial matter for two main reasons: 1)
the SCUBA beamsize at 850\mum\ (the optimal wavelength for these
studies) is ${\sim}\,15$ arcsec, with pointing errors for the telescope of
order 2$\,$arcsec; 2) the large, negative K-corrections
of dusty star-forming galaxies at these wavelengths (i.e.~the increase
in flux density as the objects are redshifted, because of the steep spectrum
of cool dust emission) imply that sub-mm observations
detect such objects at $z\,{>}\,1$ in an almost distance-independent
manner (Blain \& Longair 1993).  Thus several candidate optical
galaxies are often present within the positional uncertainty of the sub-mm
detection, and there is a very strong possibility that the actual
counterpart is at much higher redshift and undetectable with current
optical imaging.  Because of these difficulties, it is presently
unknown whether these newly discovered sub-mm galaxies are drawn from
a population similar to the LBGs, or whether the two methods select
entirely different types of object at similar redshifts.

Attention has so far focussed on constraining SCUBA source counts and using
follow-up observations in other wavebands to identify the sub-mm galaxies.
Direct estimates of the source counts have recently come from
several survey projects covering small regions on the sky 
to 3$\sigma$ sensitivities ranging from $0.5\,$mJy to $8\,$mJy 
(Smail et al.~1997;
Hughes et al.~1998; Barger et al.~1998; Holland et al.~1998;
Eales et al.~1999; Blain et al.~1999b; Chapman et al.~1999b).
In addition, statistical upper limits can  be put on the source counts at
even fainter limits through fluctuation analysis of `blank field' data
\cite{Hughes,BorCS}.
There is currently some debate about what fraction of detected sources lie
at redshifts above $z\,{\sim}\,2$, and what fraction may be at more modest
distances,
$z<1$ (see Hughes et al.~1998; Smail et al.~1998; Lilly et al.~1999; Barger
et al.~1999).
It is fair to say that the numbers are still so small, and the optical
identification procedure still sufficiently ambiguous, that this debate
is currently unresolved.
Furthermore, the issue of the importance of AGN-fuelled
star formation, i.e.~the fraction of SCUBA-bright sources which are active
galaxies, is also still an open issue (see e.g.~Ivison et al.~1999a;
Almaini et al.~1999).

A statistical estimate of the number density of LBGs whose UV spectra
imply SFR $>400\,{\rm M}_\odot\,{\rm yr}^{-1}$, after correcting for extinction
and including photometric errors (Adelberger \& Steidel~2000),
results in a comoving number density that is
roughly comparable to the comoving number density of SCUBA sources
with this SFR
(2000 per square degree -- e.g.~Eales et al.~1999; Blain et al.~1999b).
This raises the question of whether the populations are the same or at
least related, and whether,
if the uncertainties were better constrained in the LBG
population, and if more near-IR data were available,
it would be relatively easy to select star forming galaxies
detectable with SCUBA.

In order to address these issues, we have targetted a sample of LBGs,
which have high UV-estimated SFRs, for photometry in the sub-mm.  The
full sample of more than 700 LBGs which have spectroscopic redshifts
(Steidel et al.~in preparation) has a range of extinction implied by
far-UV models from zero to a factor of more than 100.  We have chosen
a small number of candidates from this group whose UV properties
indicate that they are the most likely to be detectable with SCUBA.

The sub-mm observations provide an estimate of the global dust mass
and SFR which is unaffected by the obscuration uncertainties inherent
in the UV continuum, but on the other hand suffers from many model
dependencies.  Relating the SFRs obtained from the two wavelength
regimes may help to elucidate the energetics of star formation in
luminous star forming galaxies.
The bulk of this paper therefore focuses on comparison of SCUBA flux
densities with UV-predicted 850\mum\ emission, or equivalently of the
SFRs estimated from the rest-frame UV and far-IR wavebands.
 
\section{Observations}
\label{sec:observations}

A total of 16 LBGs were targetted in three different regions: the HDF
flanking fields \cite{HDF}, the Westphal 14\,hour field (also known as
the `Groth Strip', Groth et al.~1994), and several deep redshift
survey fields at 22 hours (Steidel et al.~in preparation).  These are
three of the regions for which extensive study of LBGs has already
been carried out (see e.g.~Steidel et al.~1996a). 

Our initial choice of targets was based on the qualitative assumption that
high UV-derived SFRs and bright magnitudes might imply large sub-mm fluxes.
Hence the LBGs chosen for sub-mm observation had the largest SFRs inferred
from the UV and optical data available at the time of observation, ranging
from 360 to $860\,{\rm M}_\odot\,{\rm yr}^{-1}$
(corresponding to 1.8--4.3\,mJy at
850\mum\,).  Subsequent careful analysis of the errors through Monte
Carlo simulations (Adelberger \& Steidel~2000) have shown that
actually the brighter galaxies, with relatively small corrections to their
UV SFRs, are more likely to have large sub-mm fluxes.  In addition,
optical spectral and imaging data, obtained
subsequent to the sub-mm observations, also tended to lead to reductions
in the initial estimates; in fact
a few of the objects which had the most extreme reddening estimates
turned out to be at much lower redshift.
As a result, half of the
original 16 LBGs targetted no longer have implied SFRs which would make
them detectable with SCUBA.  Nevertheless, it is still interesting to
study carefully what the observed sample may be telling us.

Those objects with implied SFRs which place them at a SCUBA flux
density level of at least 0.5\,mJy are included in Table~1. (In fact
all have implied 850\mum\ flux densities $>$1\,mJy).  The remaining
objects are included in Table~2, and they can be regarded, in a sense, as
a control sample.  There are no redshift estimates for the objects in
Table~2.  Object names in the tables denote
catalogue entries for these fields as discussed in Steidel et al.~(in
preparation).

\subsection{Submillimetre observations}

The observations were conducted with the Submillimetre Common-User
Bolometer Array (SCUBA, Holland et al.~1999) on the James Clerk
Maxwell Telescope.  The data set was obtained over six nights from
observing runs in May and June 1998. We operated the 91 element
Short-wave array at 450\mum\ and the 37 element Long-wave array at
850\mum\ simultaneously in photometric mode, giving half-power beam
widths of 7.5, and 14.7 arcsec respectively.

Conditions were generally reasonably good in May, with $\tau_{850}$ ranging
from 0.31 to 0.40, and very good in June ($\tau_{850}$=0.13--0.16).
Observations were divided into scans lasting about $40\,$min for
$100\times18\,$s integrations.  The usual 9-point jiggle pattern
was employed to reduce the impact of pointing errors and thereby
produce greater photometric accuracy by averaging the source signal
over a slightly larger area than the beam.  Whilst jiggling, the
secondary was chopped by 45 arcsec in azimuth at the standard
7.8125\,Hz.  This mode allows deeper integration on the central pixel
for a fixed observing time than in mapping mode.  Pointing errors were
below 2 arcsec, checked on nearby blazars every $80\,$min.  The total
exposure times and resulting rms sensitivities are listed in Tables~1
and 2.

The data were reduced using both the Starlink package SURF (Scuba User
Reduction Facility, Jenness \& Lightfoot~1998), and independently with
our own routines (see Borys, Chapman \& Scott~1999).  Spikes were
rejected from the double difference data, which were then corrected for
atmospheric opacity and sky emission using the median of all the
pixels except for the central pixel and any obviously bad pixels.  We
improved the effective sensitivity to source detection by incorporating
the flux density in negative off-beam pixels of the LBG source, as
described in the following sub-section, and by removing possible
sources in the field before estimating background flux levels; any
bolometer that had a signal $> 2 \sigma$ was removed from the sky
estimation.  Whether or not the negative beam bolometers were used in the
sky subtraction made no discernible difference.

%
% Table 1
%
\begin{table*}
\begin{center}
\caption{Sub-mm flux densities and other data for our targetted Lyman
break galaxies
(and those already covered by the Hughes et al.~1998 HDF integration),
including only those objects with relatively high implied star formation rates.
We give the galaxy designation, redshift, 
SCUBA integration time (total time chopping on and off source),
observed 850\mum\ and 450\mum\ flux density and error, and
the SCUBA-derived SFR upper
limit in the first six columns.  
Columns 7 through 10 provide the associated optical
parameters: $R$-band magnitude, the SFR derived from the raw UV flux and the 
UV-corrected SFR estimate, plus
the 850\mum\ flux density predicted from the
UV data.  SFR estimates assume $\Omega_0=1$ and $h=0.5$.}
\begin{tabular}{@{\extracolsep{-1.5pt}}lcccccccccc}
\noalign{\medskip}
\hline
{Galaxy} & {$z$} & \multispan4{\hfil Sub-mm parameters \hfil } &
	\multispan5{\hfil Restframe UV parameters \hfil } \cr 
{} & {} & $t$ & $\!S_{850}\pm\sigma_{850}\!$ & 
	$\!S_{450}\pm\sigma_{450}\!$ & %L$_{(bol,850)}$ &
SFR$_{850}^{\rm \,a}$ &
{$R_{\rm s}$} & $G\,{-}\,R_{\rm s}$& % L$_{(bol,UV)}$ &
SFR$_{\rm UV}^{\rm \,b}$ & $\!\!{\rm SFR}_{\rm UV\!-corr}^{\rm \,c}\!\!$ & 
$S_{\rm 850-UV}^{\rm \,d}$ \cr
{} & {} & (ksec) & (mJy) & (mJy)& $\!\!({\rm M}_\odot\,{\rm yr}^{-1})\!\!$&
 (mag) & (mag) & (${\rm M}_\odot\,{\rm yr}^{-1}$)&
 $\!({\rm M}_\odot\,{\rm yr}^{-1})\!$& (mJy) \\
\noalign{\medskip}
\noalign{{\bf WESTPHAL FIELD}}
\noalign{\medskip}
W-DD20 & 3.083& 9.0& $-0.1\pm0.9$& $-11\pm10\phe$ & ${<}\,190$& 23.1&
 1.2& 51.2&  193$\pm$210 & 1.6$\pm$1.8  \\ % Td=38K 4.17$\pm$ 4.76
W-MM27 & 2.789& 9.0& $-1.1\pm1.0$& $\phm35\pm12^{\rm e}$& ${<}\,120$& 24.1&
 1.0& 20.5&  175$\pm$164 & 2.2$\pm$1.5 \\ %5.77 +-  3.91\\
W-MM8  & 2.829& 5.4& $\phm0.7\pm1.1$& $\pho{-}8\pm16\phe$ & ${<}\,250$& 24.1&
 1.0& 20.4&  175$\pm$164 & 1.5$\pm$1.1 \\ %3.94 +-  2.86
W-MMD109& 2.715& 9.0& $\phm1.4\pm0.9$& $\pho{-}7\pm15\phe$ & ${<}\,250$& 23.9&
 0.8& 22.5&  280$\pm$173 & 1.8$\pm$1.3 \\ %4.90 +-  3.42 
W-CC76 & 2.871& 1.8& $-1.7\pm2.5$& $\phm47\pm54\phe$ & ${<}\,390$& 23.3&
 0.8& 37.9&  201$\pm$124 & 1.4$\pm$0.8 \\ %3.60 +-  2.22
W-CC1 & 2.984& 5.4&  $\phm0.5\pm1.3$& $\pho\phm1\pm20\phe$ & ${<}\,310$& 23.8&
 1.0& 25.5&  152$\pm$187 & 1.7$\pm$1.4 \\ %4.56 +-  3.73
W-MMD11 & 2.979& 4.9&  $\phm5.5\pm1.4$& $\phm22\pm23\phe$ & $600\pm150$& 24.1&
 1.0& 21.0&  173$\pm$95 & 1.2$\pm$0.6  \\ %3.10 +-  1.70
W-MMD46 & 2.917& 4.9& $-0.6\pm1.1$& $\phm14\pm16\phe$ & ${<}\,190$& 23.8&
 0.9& 26.7& 138$\pm$92 & 1.6$\pm$1.0 \\ %4.12 +-  2.73
\noalign{\smallskip}
{\bf MEAN}$^{\rm \, f}$ & & &${\bf \phm0.51\pm0.39}$ &
 ${\bf \phm4.9\pm5.3\phe}$ & & & & & & ${\bf 1.46\pm0.36}$ \\
\multispan 3{\bf MEAN w/o W-MMD11 \hfil }&${\bf \phm0.08\pm0.41}$ &
 ${\bf \phm3.9\pm5.5\phe}$ & & & & & & ${\bf 1.61\pm0.44}$ \\
\noalign{\medskip}
\noalign{{\bf HDF (Hughes et al.~1998)}}
\noalign{\medskip}
H-MM18 & 2.929& 135$^{\rm \,g}$& $<1.2^{\rm \,h}$& ${<}\,20$& ${<}\,120$& 24.1&
 1.0& 20.0& 151$\pm$96 & $1.1\pm0.8$ \\
H-MM17 & 2.931& 135$\phe$& $<1.2\phe$ & ${<}\,20$& ${<}\,120$& 24.5&
 1.0& 14.4& 121$\pm$86 & $0.7\pm1.0$ \\

\hline
\end{tabular}
\end{center}
\begin{flushleft}
{$^{\rm a}$ For the upper limits we quote the Bayesian 95 per cent value
(i.e.~excluding the unphysical negative flux region),
assuming a 50\,K dust temperature, as described in section~3.}\\
{$^{\rm b}$ Uncorrected, based on the UV flux.}\\
{$^{\rm c}$ Calzetti attenuation curve corrected, taking into account
Monte Carlo simulations of the photometric errors.  Simple rms errors
are quoted, while in fact the distributions are skewed, and none of the
predicted SFRs or flux densities could be negative.  The order of the
LBGs in this list reflects our original estimates for SFR.}\\
{$^{\rm d}$ Predicted from the UV colours, see section~3.}\\
{$^{\rm e}$ This possible detection at 450\mum\ is much more likely to
be from a foreground object than the $z\,{\sim}\,3$ galaxy.}\\
{$^{\rm f}$ Average flux from all targets, combined with inverse variance
weighting.}\\
{$^{\rm g}$ Approximate 2$\sigma$ upper limits estimated from
Hughes et al.~(1998).}\\
{$^{\rm h}$ The 51 hour integration on the HDF was done in `jiggle map'
mode to fully sample the field; it is roughly equivalent to a 10 hour
`photometry' integration.}
\end{flushleft}
\end{table*}

%
% TABLE 2
%
\begin{table*}
\begin{center}
\caption{Sub-mm flux densities and other data from Lyman break galaxies
originally on our target list, but which subsequently turned out to have
significantly lower or uncertain estimates for star formation rate than those
in Table~1.  This can be considered as a control list, and we present the
data for completeness. 
The objects in the various 22 hour fields were originally in our sample
because they had extreme reddening estimates, but subsequently have turned
out not to be LBGs, and probably lie at lower redshifts. They are not likely
to be sub-mm emitters.  Columns are as in Table~1.}
\begin{tabular}{@{\extracolsep{-1.5pt}}lcccccccccc}
\noalign{\medskip}
\hline
{Galaxy} & {$z$} & \multispan4{\hfil Sub-mm parameters \hfil } &
        \multispan5{\hfil Restframe UV parameters \hfil } \\
{} & {} & $t$ & $\!S_{850}\pm\sigma_{850}\!$ &
        $\!S_{450}\pm\sigma_{450}\!$ & %L$_{(bol,850)}$ &
SFR$_{850}^{\rm \,a}$& {$R_{\rm s}$}& $G\,{-}\,R_{\rm s}$& % L$_{(bol,UV)}$&
SFR$_{\rm UV}^{\rm \,b}$ & $\!\!{\rm SFR}_{\rm UV\!-corr}^{\rm \,c}\!\!$&
$S_{\rm 850-UV}^{\rm \,d}$ \\
{} & {} & (ksec) & (mJy) & (mJy)& $\!\!({\rm M}_\odot\,{\rm yr}^{-1})\!\!$&
 (mag) & (mag) & (${\rm M}_\odot\,{\rm yr}^{-1}$)&
 $\!({\rm M}_\odot\,{\rm yr}^{-1})\!$& (mJy) \\
\noalign{\medskip}
\noalign{{\bf HDF FLANKING FIELDS}}
\noalign{\medskip}
H-C38  & 3.114& 3.6& $\phm0.6\pm1.9\pho$& $-144\pm81\pho$ &  ${<}\,490$& 25.0&
 0.7& 5.0& $31.0\pm28.0$ &  $0.14\pm0.12$\\
H-C10  & 2.985& 3.6& $-1.7\pm1.8\pho$& $\phm129\pm67\pho$ &  ${<}\,280$& 24.8&
 0.8& 6.1& $44.0\pm20.1$ & $0.10\pm0.11$ \\
H-D2   & 2.982& 1.8& $-5.0\pm3.1\pho$& $-162\pm107$ &  ${<}\,400$& 25.2&
 0.7& 0.4& $2.1\pm1.0$ & $0.01\pm0.04$\\
\noalign{\medskip}
\noalign{{\bf 22 HOUR FIELDS}}
\noalign{\medskip}
22A-MD6& --- & 7.2&  $\phm0.1\pm1.1\pho$& $\pho\pho{-}6\pm16\pho$ & --- & 24.1&
 0.9& --- & --- & --- \\
CDFA-M17& ---& 5.4& $-0.7\pm1.1\pho$& $\phm\pho10\pm14\pho$ & --- & 24.6&
 1.0& --- & --- & --- \\
22A-MD55&--- & 5.4& $-1.9\pm1.1\pho$& $\phm\pho21\pm12\pho$ & --- & 24.4&
 0.9& --- & --- & --- \\
DSF22A-D2&---& 3.6&  $\phm1.0\pm1.2\pho$& $\pho\pho{-}1\pm13\pho$& ---& 23.8&
 0.7& --- & --- & --- \\
22B-MD36& ---& 3.6& $\phm0.0\pm1.2\pho$& $\pho\pho{-}2\pm12\pho$ & ---& 23.2&
 0.8& --- & --- & --- \\
\noalign{\smallskip}
{\bf MEAN}$^{\rm \, f}$ & & &${\bf -0.51\pm0.46\pho}$ & ${\bf \phm4.9\pm5.8}$&
 & & & & & ${\bf 0.03\pm0.04}$ \\
\hline
\end{tabular}
\end{center}
\end{table*}

We have written routines which extend the capabilities of SURF by
analysing the photometry of each bolometer independently, weighting
each scan by its variance. This is crucial for data of the same object
taken over several nights of observing.  Our routines also allow us to
characterise the noise spectrum and to check for residual gradients
across the array.  In practice, however, we found it was unnecessary
to subtract gradients from this data set.
Analysis of the entire array
used in photometry mode is essential to extract meaningful numbers
for faint signal levels, since we are approaching the noise levels
for which source counts are beginning to make a sizable contribution
to the background noise. Confusion limit is at ${\sim}\,1\,$mJy (Hughes
et al.~1998).  We discuss this further in Section 2.1.2.

We calibrated our data against Uranus and IRC\,10216. The calibrations
agree with each other, and with the gains found by other observers
using SCUBA at around the same time, to within 10 per cent at 850\mum, and to
20 per cent at 450\mum, remaining stable night to night.  Folding this
calibration uncertainty in to our error budget has little effect
at the low signal-to-noise ratio levels of these observations.

%___________________________________________________________________________
\setcounter{figure}{0}
%%%%% FIGURE  array  (1) %%%%%%
%
\begin{figure}
\begin{center}
\psfig{file=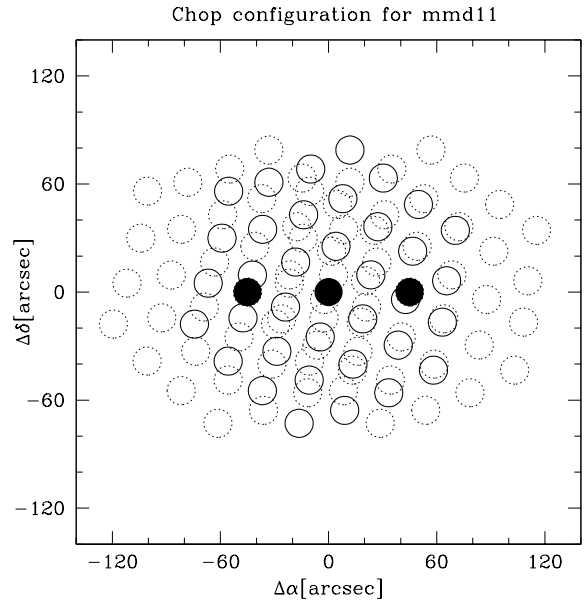,height=8cm}
\caption{Chopping geometry.  The SCUBA 850\mum\ bolometer configuration
on the sky is depicted.  The solid filled circles give the
position of the central bolometer in all three chop positions (the chop
throw for these observations was 45 arcsec).  The open circles with solid 
lines show the array in the central chop position, while the dashed 
circles give the off-beams.  Clearly there will be
some off-beams that chop on to
the source, with an efficiency which varies as the array rotates relative
to the sky.}
\label{fig:array}
\end{center}
\end{figure}
%___________________________________________________________________________

\subsubsection{Folding in the off-beams}

The signal from the off-beams that chop on to the source can be used to
improve the  estimate of the flux density from the source.  For a source with 
flux density $S$, measured with an efficiency $\epsilon$ and measurement
error $\sigma$, the probability that the measured value is $x$ is given by:
\begin{equation}
\label{equ:prob}
P(x) \propto \exp\left[-{{1}\over{2}}\left({{x - \epsilon S}
 \over{\sigma}}\right)^2\right].
\end{equation}
By minimizing the joint probability of $N$ measurements, the maximum likelihood
estimator of the source flux density is
\begin{equation}
\label{equ:max}
\bar{S} = {{\sum_i^N {x^\prime}_i {\sigma^\prime}_i^{-2}}\over{\sum_i^N
{\sigma^\prime}_i^{-2}}},
\end{equation}
where $x^\prime = x/\epsilon$ and $\sigma^\prime = \sigma/\epsilon$.

As the sky rotates during an experiment, the effective efficiency,
$\epsilon$, for the off beams varies.  Fig.~1 shows an
illustration of this effect during our observations of W-MMD11.
For our double-difference observations there are instantaneously
$N\,{=}\,3$ beams, with the
central beam having an efficiency of unity and the two off beams having
\begin{equation}
\epsilon=-0.5\exp\left(-{d^2\over2\sigma_{\rm b}^2}\right),
\end{equation}
where $d$ is the angular distance of the off-beam centre from the source, and
$\sigma_{\rm b}$ is the Gaussian width of the beam.

In the case of W-MMD11, our detection level increases from
${\simeq}\,3.0\sigma$ to 3.9$\sigma$, after folding in the negative flux
density from the outer pixels.

Our sub-mm flux density measurements for all of our targets
are presented in Tables~1 and 2, including the upper limits at 450\mum\
as well as the 850\mum\ data.

\subsubsection{Confusion noise}
Given that our 850\mum\ integrations go relatively faint, we need to
be concerned about the issue of confusion
noise \cite{Scheuer57,Scheuer74,Condon,Walletal}.  In other words, we need
to consider to what extent the fluctuations due to undetected sources
contribute to our error bars.  This is additionally complicated by our
use of double-difference data, rather than data from fully-sampled maps.

Blain et al.~(1998) quote a value of $0.44\,$mJy for the variance due
to confusion, derived from their source counts.
We find that any reasonable fit to the counts,
including extrapolation to low fluxes (with the
constraint of not over-producing the far-IR background) lead to a
variance of no more than $\sigma_{conf}\,{\simeq}\,0.5\,$mJy.
This then is the confusion noise for a single bolomter observation of
the sub-mm sky.  Since the JCMT gives a triple-beam response,
i.e.~${\rm On}-({\rm Off}_1+{\rm Off}_2)/2$,
then the rms in our photometry observations is expected to
be $\sqrt{3/2}$ higher than this, for the simplest configuration.  But,
since we chopped in azimuth while the sky rotated, then in practice
we are taking the central value minus the average of several other sky
positions.  So the rms of
our photometry observations ends up being only a little higher than
for an individual bolometer measurement.  We checked with Monte Carlo
simulations (of sources drawn randomly from reasonable count models)
that the confusion noise for our observations was unlikely to be higher than
$0.55\,mJy$.  This represents the expected error on our fluxes due to
the presence of undetected sources in the beam.  We confirmed this
value with Monte Carlo simulations of a large number of sources drawn
from a model of the number counts.

This confusion noise is already contained in our error bars, and should
not be added to the error budget.  
Our noise estimates come from the variance among the different
flux samples of each target.  Therefore our noise estimates will
already contain at least part of the confusion noise, since
the off-beams were sampling different positions on the sky.  Hence
the variation in the flux of an object throughout the observation
period could have partly been due to confusion noise.  We believe that
the error bars we quote on the flux for each target is a reasonable
estimate of the uncertainty in the flux at that central sky position.

Each of our target measurements had a final estimated uncertainty of
about $1\,$mJy.  Since confusion and noise and noise from the atmosphere
or the instrument add in quadrature, then the contribution from
confusion noise is sub-dominant -- contributing less than
30\% to the variance.
Note also that when we average our measurements together (in Section~4.1)
we can reach, in principle, below both the individual sky/instrumental
noise and the confusion noise in each measurement.

%___________________________________________________________________________
%
% FIGURE keckspectra  (2)
%
\begin{figure*}
\begin{center}
\psfig{file=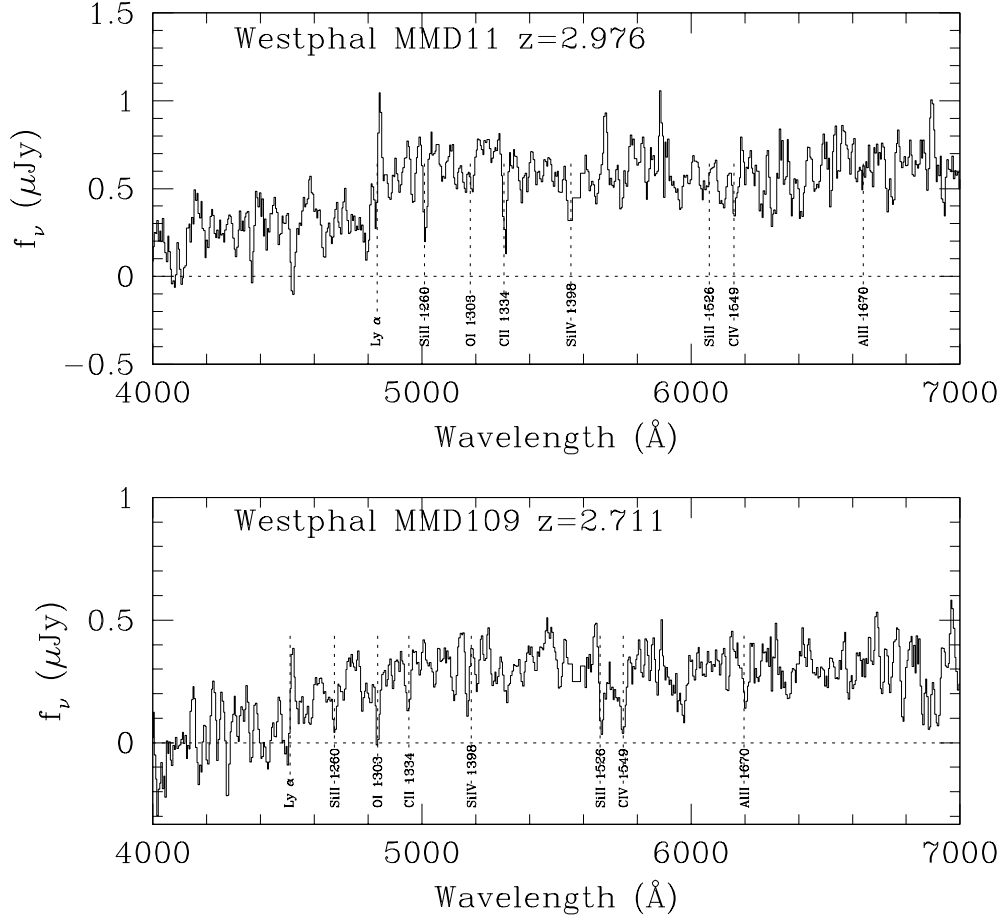,height=16cm}
\caption{Optical spectra for two representative Lyman break galaxies, obtained
using LRIS on Keck.
The redshifted wavelengths of Ly$\,\alpha$ emission and some prominent
interstellar absorption lines are marked.  Note that the vertical axis
depicts only relative flux.}
\label{fig:keckspectra}
\end{center}
\end{figure*}
%___________________________________________________________________________

\subsection{Optical spectroscopy}

Spectra for these objects were obtained using the LRIS instrument on
the Keck telescope in the course of following up the $U$-band dropout
candidate objects in the survey fields, to confirm their redshifts.
Details of the observations can be found elsewhere
\cite{Ste96b,adelberger00}.  Spectra typically show weak Ly$\,\alpha$
emission and several absorption lines from the interstellar medium
within these galaxies.  None of our targetted objects have remarkable
optical spectra, strong emission lines being absent for example.
Fig.~\ref{fig:keckspectra} shows the spectra of
two representative objects, including the possible sub-mm detection.

\section{Comparing the UV with the far-IR}

In order to compare optical (rest-frame UV)
observations to our sub-mm (rest-frame far-IR) data we can
calculate a far-IR flux consistent with the UV extinction and compare
that to our direct SCUBA measurements.
Alternatively, we can
calculate the SFR implied by each of the data sets (UV and sub-mm)
and compare those.
%For the weakest sources only
%this second comparison is made.  We often measure a negative flux
%value and hesitate to infer either a negative dust temperature or
%negative SFR from it.
For our one detected source, the second, perhaps less direct comparison
may help elucidate the underlying physical processes.
We list both sets of estimates in Table~1, and now describe how we
arrived at the values given there.

\subsection{Estimates of star formation rates from the sub-mm data}

To estimate the SFR from the measured sub-mm flux density we follow
an approach which has become
conventional for dealing with {\sl IRAS} galaxies, for example.
In essence this involves estimating the $60\,\mu$m flux, which is
approximately linearly related to the SFR.

We first calculate the quantity
\begin{equation}
{\cal L}_{\rm rest}\equiv
 \left(\nu L_\nu\right)_{\rm rest} = 4\pi D_{\rm L}^2 \nu_{\rm obs} S_{\rm obs},
\end{equation}
where $L_\nu$ is the luminosity density and $D_{\rm L}$ is the usual
luminosity distance.  In the rest frame we are observing the galaxy at
$850/(1+z)\,\mu$m.  We then assume that the dust can be described by a
mass absorption coefficient of the form
\begin{equation}
k_{\rm d} = 0.14\, (\lambda/850\,\mu{\rm m})^{-\beta_{\rm d}}
 \,{\rm m}^2{\rm kg}^{-1}
\end{equation}
(see Hughes et al.~1997 and references therein), with a typical value of
$\beta_{\rm d}=1.5$ for the index.  Note that other authors
(e.g.~Lisenfeld, Isaak \& Hills~2000 and references therein) suggest values
of $k_{\rm d}$ which are smaller by a factor ${\sim}\,1.5$; our
dust masses (and other derived qunatities) are thus fairly conservative.

Since
\begin{equation}
L_\nu\propto S_\nu \propto k_{{\rm d},\nu}B_\nu,
\end{equation}
with
$B_\nu(T_{\rm d})$ the Planck function, then we can obtain
${\cal L}(60\,\mu{\rm m})$ from ${\cal L}({\rm rest}\!)$ using the
ratio of $\nu k_{{\rm d},\nu}B_\nu$ at the two wavelengths.
We also assume for definiteness that
$H_0\,{=}\,50\,{\rm km}\,{\rm s}^{-1}{\rm Mpc}^{-1}$
and $\Omega_0\,{=}\,1.0$.  Both $L_{\rm FIR}$ and SFR (as well as inferred
properties such as dust mass)
scale as $h^{-2}$, and at these redshifts, the values are about 2
times higher in an open $\Omega_0=0.3$ universe.
 
As an intermediate step we could also estimate the dust mass at this point,
assuming that the dust is optically thin.  The dust mass estimate is just
\begin{equation}
M_{\rm d} = \frac{S_{\rm obs} D_{\rm L}^2}
 {k_{\rm d}^{\rm rest} B(\nu^{\rm rest},T_{\rm d})\,(1+z)}
\label{equ:dustmass}
\end{equation}
(Hughes et al.~1997),
with $B(\nu^{\rm rest},T_{\rm d})$ the Planck function evaluated in the
rest frame, assuming the dust to have a temperature $T_{\rm d}$.  We take
$T_{\rm d}\,{=}\,50\,$K as our standard value.
Note that we are assuming that
the full sub-mm flux is due to thermal dust emission.
A strong synchrotron flux would cause our mass estimates to be high.
Furthermore if the dust is not entirely optically thin this would also
affect the estimate of $M_{\rm d}$ and correspondingly ${\cal L}$.

We can proceed from the estimate of ${\cal L}(60\,\mu{\rm m})$ to the
SFR by firstly applying a bolometric correction, and then using a
standard conversion factor.  Rowan-Robinson et al.~(1997) suggest that
$L_{\rm FIR}\,{\simeq}\,1.7\,{\cal L}(60\,\mu{\rm m})$,
where $L_{\rm FIR}$ means the total luminosity over say
$1\,\mu$m--$1000\,\mu$m.  The far-IR luminosity is expected to be directly
proportional to the star formation rate ${\rm SFR}$, i.e.
\begin{equation}
L_{\rm FIR} = K\times{\rm SFR}, 
\label{equ:LFIRSFR}
\end{equation}
where the coefficient $K$ is estimated from well-studied local
objects to be $K=2.2\times\,10^{9}\,{\rm L}_\odot\,{\rm M}_\odot^{-1}\,$yr
(Rowan-Robinson et al.~1997).  
Estimates in the literature range from
$1.5\times10^{9}$ to
$4.2\times10^{9}\,{\rm L}_\odot\,{\rm M}_\odot^{-1}$\,yr
(Thronson \& Telesco 1986; Scoville \& Young 1983).
Significant differences between local and distant galaxies could of
course change this scaling factor.  

Note the limitations of this procedure in assuming an isothermal distribution;
dust components with higher temperature, or AGN contributions to the flux,
could lead to higher $L_{\rm FIR}$ for the same $M_{\rm d}$.
The estimated far-IR flux and SFR will also strongly depend on the
assumed form of the grey body.  Bolometric corrections will typically
vary as $T_{\rm d}^{4+\beta_{\rm d}}$ (see e.g.~Blain et al.~1999c)
and so different dust temperatures
or emissivity indices can give significantly different results.

We make an empirical estimate of the total uncertainty involved in the dust
models by fitting Eq.~(6), for $T_{\rm d}\,{=}\,50\,$K, to
the long wavelength data for several vigorous star-forming galaxies
for which we have rest frame IR data.  The discrepancy between our
fits at $60\,\mu$m and the actual data at
$\lambda=850\,\mu{\rm m}/(1+z)\,{\sim}\,200\,\mu{\rm m}$ spans a factor
of 3.  We adopt this as the uncertainty in $L_{\rm FIR}$ inferred from our
SCUBA data (see Hughes \& Dunlop~1998 for a recent discussion of the various
uncertainties in such estimates).

\begin{figure}
\begin{center}
\psfig{file=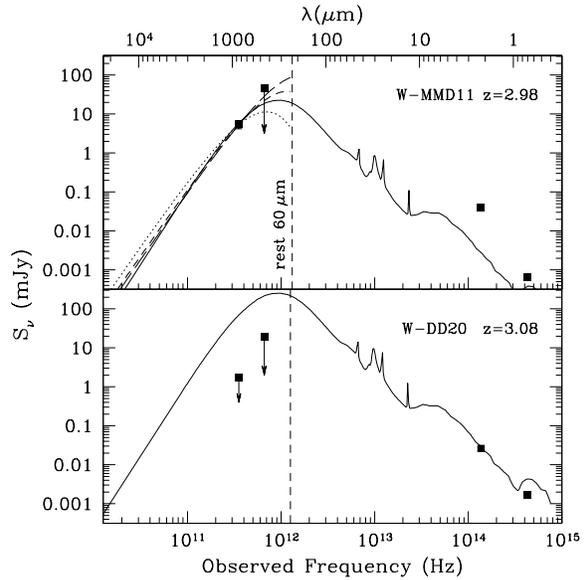,width=8cm}
\caption{The measured $S_{850}$, $S_{450}$, $K$-band and $R_{\rm s}$-band
points and 2$\sigma$ upper limits are plotted
for our detection (W-MMD11, upper panel) and also for 
our target LBG with the highest achieved sensitivity (W-DD20, lower panel).
The solid line spectrum shown here is a redshifted version of a
nearby star-forming galaxy, M82, adapted from
Efstathiou et al.~(1999), and normalized to the 850\mum\
point (upper panel) and the near-IR data (lower panel).
In the case of W-DD20, the
normalization to the optical data points emphasizes the poor fit of this
local starburst to the majority of our LBG sample.
The dashed lines in the upper panel
are greybody spectra for emissivity $\beta_{\rm d}=1.5$ 
and $T_{\rm d}\,{=}\,30$, 50 and 70\,K.
For nearby galaxies, the IRAS 60\mum\ and 100\mum\ bands have 
typically been used to estimate SFRs (e.g.~Meurer et al.~1997).
We have marked the location of the 60\mum\
band, in the rest-frame, with a vertical dotted line for both objects, as an
indication of the extrapolation used in this study relative to nearby objects.
The spread of values of the dust curves
at $60\,\mu{\rm m}$ indicates the level of uncertainties in our dust
model.}
\label{fig:seds}
\end{center}
\end{figure}

Fig.~\ref{fig:seds} provides an illustration of the uncertainties
involved in estimating $L_{\rm FIR}$ from measurements at a single
wavelength.  
Dashed curves in Fig.~\ref{fig:seds} are
for model dust emission spectra with $T_{\rm d}$ = 30, 50,
and 70\,K, normalized to our measurement.
Notice that these three curves differ by a factor of three at
$\lambda=60\,\mu{\rm m}$, the wavelength at which one ordinarily infers
the SFR for nearby galaxies.  In principle we could use our own
$450\,\mu{\rm m}$ data to remove the uncertainty in dust parameters, but
none of our data are precise enough to be very useful. 
For W-MMD11 we can infer that $T_{\rm d}\la90\,$K, with much weaker limits
for the other objects.

The data for our one detected source, W-MMD11, and for our
highest sensitivity non-detection, W-DD20, are shown in Fig.~\ref{fig:seds},
along with the redshifted spectrum of a representative star-forming
galaxy, M82 (using the model fit from Efstathiou, Rowan-Robinson
\& Siebenmorgen~1999).
W-MMD11 is fairly well
approximated by the M82 SED, except for the larger $K$-band flux.
W-DD20, on the other hand, has a much lower sub-mm flux than M82, when
normalized to the $K$ and $R_{\rm s}$ magnitudes; DD20 is clearly not an
analogue of nearby starburst galaxies.

%_____________________________________________________________
%
% FIGURE  seds  (3)
%_____________________________________________________________

\subsection{Estimate of the sub-mm flux density directly from the UV}
\label{sec:uvestimate}

To estimate the $850\,\mu$m flux, $S_{850-{\rm UV}}$ we use the
Meurer et al.~(1999) empirical relationship
between UV slope $\beta$ and the ratio
$L_{\rm FIR}/{\cal L}(1600\,{\rm\AA})$, where
${\cal L}(1600\,{\rm\AA})$ is $\nu L_\nu$
at $1600$\AA\ and $L_{\rm FIR}$ the approximately bolometric dust luminosity
as estimated from {\sl IRAS} 60 and $100\,\mu$m fluxes.
{}From this we can directly estimate the flux at $850\,\mu$m, assuming
an isothermal modified blackbody with a dust temperature of $50\,$K and an
emissivity index of 1.5 (see for example Rowan-Robinson et al.~1990).
This approach is consistent
with the Calzetti law applied to observations of local starburst
galaxies (Meurer et al.~1997; Calzetti et al.~1996).
Ouchi et al.~(1999) have also demonstrated that such a procedure is
consistent with the empirical relation for $L_{\rm FIR}/L_{\rm UV}$
discussed by Meurer et al.~(1997, 1999). Our UV-based predictions for sub-mm
emission are listed in Table~1.
Note, that the
value of $K$ (from equation~(8)) that we use is approximately a factor of
3 {\it lower\/} than the effective value used by Meurer et al.~(1999).
Our results are therefore conservative in the sense that other reasonable
estimates could predict even higher sub-mm fluxes, and hence make our
non-detections even harder to understand.

\subsection{Estimates of star formation rates from the UV spectra}

It is worth describing the UV-based SFR estimates in more detail, to reveal
various sources of potential uncertainty.
The star formation rate is estimated from the rest-frame UV by
calculating an implied bolometric luminosity ($L_{\rm bol}$) and
extrapolating to the SFR using an assumed initial mass function (IMF).
Dust opacity leads to both a global dimming and a reddening
of a galaxy's SED, with the UV region being most affected.  
We also need to take into account inter-galactic dimming, due to
intervening Ly$\,\alpha$ blanketing and
photoelectric absorption (Madau et al.~1996).
The largest uncertainty in the SFR calculation from the
rest frame UV results from the dust extinction.  Extinction curves
show some environmental dependence, e.g.~differences between the Milky
Way, the LMC, the SMC and M31 (see Calzetti~1997 for a discussion).
With extended objects the effective obscuration is a function of the
dust distribution and geometry (Calzetti et al.~1996).  We have focussed
on the dust effects modelled by Calzetti (1997), which will be
referred to hereafter as the Calzetti attenuation law (see also
Calzetti et al.~1995, 1996; Calzetti \& Heckman~1999; Meurer, Heckman
\& Calzetti~1999).  Meurer et al.~(1999) found that their general procedure
produced results which were in good agreement with radio data (Richards
2000).

The correction factors used to obtain the UV-corrected SFRs in Tables~1
and 2 are based on the Calzetti attenuation curve.  The slope of the UV
continuum, usually referred to as $\beta$, can be estimated for
$z\,{\sim}\,3$ objects from broad band colours covering the $G$- and
$R_{\rm s}$-bands \cite{Ste96a}.  An unreddened UV stellar continuum slope of
$\beta_0\,{=}\,-2.1$, thought to be appropriate for high redshift stellar
populations \cite{Caletal95}, is used throughout; the difference
between $2.1$ and the measured slope indicates the total dust column
density.  The mean dust correction for the entire LBG sample is
${\sim}\,4.5$.  The extinction corrections are computed using a $10^9$ yr
starburst, which is likely to be a conservative assumption since
younger starbursts would give bluer intrinsic UV spectra, implying
larger reddening corrections and more dust.

The SFR can vary considerably for differences in the assumed reddening
curve.  The Calzetti model results in larger corrections, by a factor
of ${\sim}\,2$, than those derived from the SMC (Bouchet et al.~1985).
However, reddening curves derived from resolved stars, such as in the SMC,
do not include scattered light along the line-of-sight, as do galactic-derived
curves.  Meurer et al.~(1999) have recently shown that under the assumption
of an SMC-type reddening curve, the far-IR/far-UV relation observed in local
UV-selected 
star-forming galaxies would not be reproduced, because of this effect.
Our corrections to the SFR are based on the properties of local starburst
galaxies (which themselves have a far-IR/far-UV flux relation containing
considerable scatter, Meurer et al.~1997).  This relationship could of course
be different at high redshift.

In summary then, we estimate the quantity ${\rm SFR}_{\rm UV\!-corr}$ as
follows.  Meurer et al.~(1999) express their empirical
$\beta\to L_{\rm FIR}/{\cal L}_{1600}$ relation also as a
$\beta \to A_{1600}$ relation,
where $A_{1600}$ is the estimated extinction in magnitudes at $1600\,$\AA.
The SFRs are calculated using the value at $1600\,$\AA\
from this relationship
to estimate the dust-corrected $1600\,$\AA\ luminosity of the LBGs, and then
using another relationship (from Madau, Pozzetti \& Dickinson~1998)
between ${\cal L}_{1600}$ and SFR to estimate the star-formation rates.
This is a somewhat different method than was used in estimating SFRs
from the SCUBA fluxes.  We chose to follow two slightly different procedures
since one is more familiar in the optical literature and the other in the
far-IR literature.  The extent to which they are different should also
indicate the level of uncertainty.
 
\subsection{Bias caused by uncertainties in the UV-estimated SFR}
\label{sec:uverrors}

There is an obvious concern that in selecting those objects with the
highest {\it implied\/} SFRs we may have preferentially selected
statistical outliers which have the biggest positive excursions from
the true SFR, which may be considerably lower.  In other words,
it is more likely that
we have overestimated the dust correction factor than underestimated
it, due to a Malmquist-type bias.  The number density of LBGs with
small correction factors (a factor ${\sim}\,4$ or so) is much higher
than the number density with large corrections.  As a result, a galaxy
which appears to require a large correction factor may actually be an
object with a small correction factor and large photometric errors.  We
have estimated the size of this effect through Monte Carlo estimates of
error bars on UV-derived values.
In detail, we estimate the intrinsic distributions of dust-obscured
luminosities and dust extinctions from our full sample.  We can then draw
mock LBGs from these distributions, calculating the true colours of the
objects, and then measuring their colours using the same procedure as for
the real LBGs. From this procedure we can estimate the likely distribution
of dust-corrected SFRs for an object with a given redshift and $G$ and
$R_{\rm s}$ magnitudes.  This accounts for Malmquist bias, photometric errors
and most other potential sources of bias in our selection.  The resulting
distributions are not symmetric, so we simply calculate the rms scatter
in the SFRs for each LBG.
These uncertainties in ${\rm SFR}_{UV\!-corr}$ and $S_{\rm 850-UV}$ are quoted
in Tables~1 and 2.  The full procedure is described in Adelberger \&
Steidel (2000).

Further, the uncertainties in the SFR correction factor scale roughly
with the magnitude of the correction (Steidel et al.~1998), and there
is some danger that the largest implied SFRs are more uncertain than
typical.  On the other hand, while it is true that the SFR correction
factors for our sample are somewhat larger than average for LBGs, in
fact the objects with the largest corrected SFRs also have higher than
average {\it uncorrected\/} SFRs; our targets do not have the steepest
UV-slopes of the whole Lyman break sample.

The errors in inferred SFR could conceivably be relatively large
compared to the variation in UV parameters -- UV luminosity and
continuum slope in particular -- especially at fainter magnitudes.
Monte Carlo simulations (Adelberger \& Steidel~2000) reveal that
photometric errors of order 0.2 mag in $G\,{-}\,R_{\rm s}$ affect the
corrected SFR by a factor of up to 3.

\section{Detection of sources}

At 850$\mu{\rm m}$, there is one likely detection in our current
sample, Westphal-MMD11.  We have tried different weightings and edits of
the data, and find that the detection is independent of the details of
the data analysis.  The
flux density estimate (the weighted average of all scans, including
contribution from off-beams) is about $4\sigma$.  

At 450$\mu{\rm m}$, we marginally detect Westphal-MM27 at $3\sigma$.
The complete absence of flux at 850$\,\mu{\rm m}$ indicates that this
is most likely a low redshift foreground object, and we do not consider it
further.

No other galaxy among the 16 we observed is detected.

\subsection{Statistical results}

\begin{figure}
\begin{center}
\psfig{file=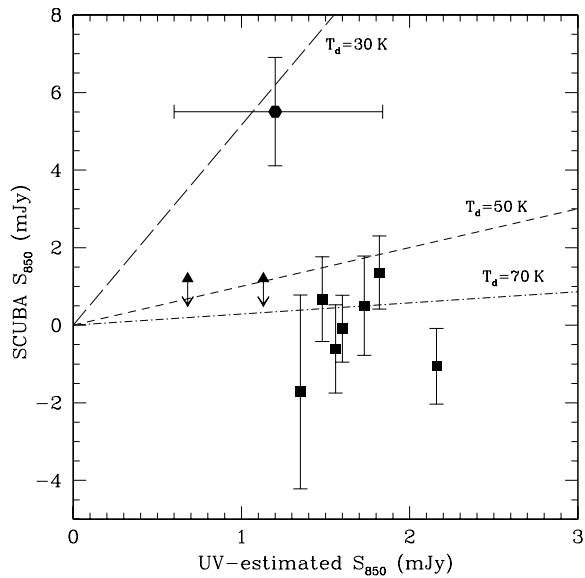,height=8cm}
\caption{ 
Measured SCUBA  850\mum\ flux densities, with $1\sigma$ error bars,
plotted against 850$\mu{\rm m}$ flux density predicted from the UV
properties and simple assumptions about star formation for
$T_{\rm d}\,{=}\,50\,$K.  Horizontal error bars are shown on our one
detection only, but should be indicative of the errors for the rest of the
points.  Although in practice the distributions are quite skewed, we have
shown the rms as a symmetric error bar for simplicity.  These errors
reflect uncertainties in photometry and other biases as estimated
using Monte Carlo simulations.
Our detection (W-MMD11) is plotted as a solid hexagon 
with 1$\sigma$ error bars.
We also show 2$\sigma$ upper limits for the 2 high SFR$_{\rm UV\!-corr}$
LBGs lying in the Hubble Deep Field (triangles).
The dashed lines show the locus around
which the points would scatter if the UV deficit is re-radiated at 
$T_{\rm d} = 30$, 50, and 70\,K.  A best-fitting
straight line to the data is consistent with zero, indicating that the 
UV-predicted flux densities are on average too high by at least a factor
of ${\sim}\,2$, as discussed in the text.
}
\label{fig:flux}
\end{center}
\end{figure}

It is important to ask whether there is statistically a detection of
the collective emission from the
eight galaxies listed in Table~1.  If we combine the
850\mum\ flux density for all the objects, with inverse variance
weighting, the mean is $0.51 \pm 0.39\,$mJy, which is consistent
with zero. Omitting W-MMD11, the result is $0.08 \pm
0.41\,$mJy.  There is little evidence that, on the average, we obtain
positive flux density when we point at our targets.  This is to be
contrasted with the total signal of $1.46 \pm 0.36$\,mJy predicted
from the UV using a dust temperature of 50\,K (as shown in
Fig.~\ref{fig:flux}), and taking into account
simulations of photometric uncertainties.  This group of LBGs thus does
not form a large part of the population of sub-mm sources.
This result points out the power of our approach (see also Scott et al.~2000):
statistical analysis of photometry allows us to measure
statistical contributions to the source counts, from targetted objects,
below $0.5\,$mJy sensitivity (around the confusion limit for mapping),
and for only a modest amount of telescope time.

In order to clarify the constraints for the sample as a whole,
in Fig.~\ref{fig:flux} we plot the actual measured sub-mm flux densities
against the predictions for $T_{\rm d}\,{=}\,50\,$K (solid squares), and
also show the predicted lines for 30, 50 and 70\,K dust models.
A best-fitting slope is consistent with zero, and certainly inconsistent with
the measured flux densities being equal to the UV-predicted ones (dashed
line labelled `50\,K', for which
$\chi^2=21.6$ for the 7 undetected objects, using the vertical
error bars).
%or 8.4, taking into account both error bars).

%___________________________________________________________________________
%
% FIGURE flux (4)
%
%___________________________________________________________________________

Another statistical approach to the data set is to ask how much lower
the SFR has to be compared with the UV-based estimates.  We constrain
the factor $f$ in the relation ${\rm SFR}_{\rm submm}=f\,{\rm SFR}_{\rm
UV\!-corr}$, by minimizing the variance for the entire sample.  We find
that $f=0.04\pm0.24$\footnote{This is using only the observational
(vertical) error bars.  If
we were to take into account the estimated horizontal error bars, we 
would obtain $f=0.07\pm0.39$.}
if we exclude the detected object, and the straight
line is a reasonable fit.
However, if we include the detection we obtain a high
value for $f$ but a terrible $\chi^2$, indicating that W-MMD11 is very
much an outlier compared with the other seven galaxies.
Excluding W-MMD11 then,
the corresponding 95 per cent upper
limit is $f<0.50$, indicating that for our undetected objects,
the sub-mm flux densities are overestimated by a factor of at least 2.  This
could be due to either a higher temperature, or one of the
other factors assumed in converting from UV to sub-mm properties.
Many of these uncertainties are dealt with in our Monte Carlo studies, as
indicated by the rms errors in Table~1.  If these (horizontal) error bars
are included in the fit, then $f=1$ cannot be very strongly excluded.

Focussing specifically on dust temperature, if we
assume that all the other parameters are held fixed,
the seven galaxies (aside from W-MMD11) in Table~1
are best described by very hot dust, at
or even above $100\,$K.  The 95 per cent
confidence lower limit for $T_{\rm d}$ is $65\,$K.

\subsection{Lyman break galaxies in the HDF}

The HDF has been mapped deeply with SCUBA \cite{Hughes}.
There are two LBGs from our catalogue lying within the Hubble Deep Field
which also have a high
UV-implied SFR (${\sim}\,150\,{\rm M}_\odot{\rm yr}^{-1}$),
corrected using the Calzetti attenuation curve.
None of the 5 sources detected
in the deep HDF SCUBA map correspond with the positions of these two LBGs.
When combined with our results (Table~1), this increases to 9 the number of
high SFR LBGs which are undetected at a $3\sigma$ level of ${\sim}\,4\,$mJy.
What we see for the HDF upper limits (Fig.~\ref{fig:flux}) is that the
UV-based predictions of the far-IR flux density continue to overestimate that
measured by SCUBA (even for objects with somewhat less extreme intrinsic
luminosities than those in our sample).  Thus the HDF non-detections
support our conjecture that the simplest UV-predicted sub-mm flux densities
are off by at least a factor ${\sim}\,2$.

\subsection{Westphal MMD11}

The first question to address is whether it is very improbable to have
found one or more ${\sim}\,4\sigma$ detections purely by chance, given
that we searched a relatively large number of objects.  The simplest
statistical estimate is that, for our full sample of 16 targets, there is
roughly a 5 per cent chance to obtain a spurious $4\sigma$ detection by
statistical fluctuation alone (2.5 per cent for our revised sample of
8 presented in Table~1).

Next we should assess the possible contamination by the sorts of galaxies
which turn up in blank field observations with SCUBA.
The probability that a random galaxy lies within the beam is given by
$P = 1 - \exp(-\pi n \theta^2)$, where $n(>S)$ is the
cumulative surface density for the population in question, and
$\theta$ is the beam radius.  For a flux density of $5.5\,$mJy the source
counts are about 1000 per square degree (e.g.~Blain et al.~1999b), and so
$P\,{\simeq}\,1$ per cent per pointing.  The chance that at least one of our
16 observations yields a $5.5\,$mJy source is then ${\sim}\,15$ per cent
(or about half of that if only our revised sample of 8 likely sub-mm
candidates is considered).
It is then statistically conceivable that the positive flux density for
W-MMD11 could be completely unassociated with the LBG, and instead due to a
random sub-mm-bright galaxy within the same SCUBA beam.  As we argue below,
circumstantial evidence for the LBG suggests that it is indeed the
source of the sub-mm emission.  But without high resolution sub-mm
observations of this region we cannot discount the interloper possibility
entirely.

There is, in fact a foreground galaxy which has slightly corrupted
the UV measurements of W-MMD11.  Fig.~\ref{fig:keckspectra}
(top panel) presents a Keck
spectrum of W-MMD11, with some of the usual lines identified.
Because of a brighter object partially on the same slit,
the continuum of W-MMD11 is over-subtracted.  The resulting spectrum then
has a slope which is inconsistent with the very red $R_{\rm s}\,{-}\,K$ colour.
For comparison the bottom panel of Fig.~2 shows the optical spectrum for
another of our targetted LBGs,
W-MMD109, which is typical of the remainder of the galaxies in the sample
(note that the vertical axis on the figure depicts only relative flux).
The spectrum of W-MMD11 is similar except for
contamination by the foreground object (4.0 arcsec North and 0.5 arcsec
East of W-MMD11), with little indication of strong emission lines for example.

W-MMD11 is, however, bright in $K$-band ($K\,{\sim}\,19.6$), and with
$R_{\rm s}\,{-}\,K=4.45$, which is about 1.5 mag redder than average for those
LBGs for which near-IR photometry has been obtained.
In fact, out of 70 LBGs for which we currently have $K$-band photometry,
only one is redder than W-MMD11.  Following recent discoveries that at
least some $z\,{>}\,2$ SCUBA sources can be identified with galaxies having
extremely red near-IR colours (Smail et al.~1999a), this argues in favour
of the identification of the SCUBA detection with the LBG, W-MMD11.
 
We discount the possibility  that the particular foreground
object is actually the sub-mm source.  This nearby source is at
$z\,{=}\,0.532$, with the [O\,{\sc ii}] line having a rest equivalent width
of $49\,$\AA, which is somewhat on the high side for general field
galaxies.  However, it has magnitudes $R_{\rm s}\,{=}\,22.16$,
$G\,{-}\,R_{\rm s}\,{=}\,0.88$ and $U_{\rm n}\,{-}\,G\,{=}\,0.54$,
indicating a normal, somewhat sub-$L^\ast$ galaxy, which
would not be expected to emit strongly at sub-mm wavelengths.  We
estimate that it could contribute at most 0.6\,mJy to the integrated
flux density within the SCUBA beam centred on W-MMD11, based on nearby
spiral galaxies \cite{Kruetal}.

The most likely explanation for the positive flux density is that we
have detected the LBG, which is re-enforced by the red $R\,{-}\,K$ of this
LBG relative to the others in our sample.  We treat it as a detection
in the discussion below.

\subsection{Comparing LBGs to sub-mm selected galaxies}

The LBG which is detected with SCUBA, W-MMD11, has a large $R_{\rm s}\,{-}\,K$
colour in addition to its extreme rest-frame UV properties.  Moreover,
only about 10 per cent of the whole current sample of LBGs (${>}\,700$
objects) are redder than W-MMD11 in terms of $G\,{-}\,R_{\rm s}$ colour,
and only 10 per cent are brighter in $R_{\rm s}$ magnitude.
It is worth exploring
how these features compare with other known high redshift SCUBA sources.

We have obtained or derived filter-matched data in the $R_{\rm s}$, $G$ and
$U_{\rm n}$ filter set (Steidel et al.~1998)
for the two high-$z$ sub-mm selected
galaxies with secure redshifts: SMM\,J02399$-$0136 (Ivison et al.~1998b) and
SMM\,J14011+0252 (Ivison et al.~1999b), with the aim of understanding whether
our selection techniques successfully predict the measured sub-mm flux
densities (explicitly, we made filter corrections for the first object,
and reobserved the second object).
Table~3 directly compares these three sub-mm bright
galaxies.  The conclusion is that SMM\,J02399$-$0136 and SMM\,J14011+0252 are
very similar to W-MMD11, although possibly more luminous (the lensing
amplification by a foreground cluster may be uncertain by
up to 50 per cent).  We also note that SMM\,J02399$-$0136 is thought to
have an AGN component (Ivison et al.~1998b), and thus would probably have
been excluded from our sample of LBGs to be studied with SCUBA.

Other LBGs with large UV-derived star-formation rates
(${>}\,100\,{\rm M}_\odot\,{\rm yr}^{-1}$),
but rather modest $R\,{-}\,K$ colours, were not
detected with SCUBA at the ${\sim}\,0.5\,$mJy rms level (when averaged
together).  This implies
that $K$-band information may help significantly to pre-select sub-mm
bright LBGs. However, photometric errors conspire to make it difficult
to know whether any particular UV-selected object is truly a
prodigious star former, at the level of the SCUBA-selected sources.

%%%%% COMPARISON TABLE SMM MMD11 %%%%%
%Table 3
\begin{table*}
\begin{center}
\caption{Comparison of the LBG detection W-MMD11, with two high-$z$
SCUBA-selected objects with confirmed redshifts, SMM\,J02399$-$0136
and SMM\,J14011+0252.}
\begin{tabular}{lccc}
\noalign{\medskip}
\hline
%\hline\hline
\noalign{\smallskip}
{Property} &{W-MMD11} &{SMM\,J02399$-$0136$^{\rm a,b}$}
  &{SMM\,J14011+0252$^{\rm c}$}\\
\noalign{\smallskip}
{$z$} & 2.979 & 2.803 & 2.550 \\
\noalign{\smallskip}
{$U_{\rm n}\,{-}\,G$} & 2.04 & 2.19 & 1.65 \\
{$G\,{-}\,R_{\rm s}$} & 1.04 & 1.00 & 0.69 \\
{$R_{\rm s}\,{-}\,K$} & 4.59 & 3.70 & 3.54 \\
\noalign{\smallskip}
{$S_{850}$ (mJy)} & 5.5$\pm$1.4 & 10.4$\pm$1.2 & 5.3$\pm$0.7\\ 
{$S_{450}$} (mJy) & 22.0$\pm$23.4 & 27.6$\pm$6.0 & 15.2$\pm$2.5 \\
\noalign{\medskip}
UV-predicted flux densities&&&\\
\noalign{\smallskip}
{$S_{\rm 850-UV}^{\rm d}$ (mJy)} & 1.2 (3.1) & 4.3 (11.1) & 8.0 (21.1)\\ 
{$S_{\rm 450-UV}^{\rm d}$ (mJy)} & 4.1 (10.6)& 13.2 (34.3)& 27.4 (51.0)\\ 
\hline
\end{tabular}
\end{center}
\begin{flushleft}
{$^{\rm a}$ Ivison et al.~1998a -- lensing factor of 2.50 assumed.}\\
{$^{\rm b}$ Note that this object has the spectrum of an AGN in the UV
and so may not conform to the far-IR/UV relation.}\\
{$^{\rm c}$ Ivison et al.~1999b -- lensing factor of 2.75 assumed.}\\
{$^{\rm d}$ Predicted from UV colours, assuming a dust temperature of 50\,K 
	(or 38\,K in brackets -- corresponding to Blain et al.~1999b).}
\end{flushleft}
\end{table*}

For the two SCUBA-selected objects in Table~3, the sub-mm flux densities and
ratio, ($S_{450}/S_{850}$), are consistent with what we would predict
from the UV magnitudes and slope (see section 3.3) using a dust
temperature of ${\sim}\,50\,$K and emissivity $\beta_{\rm d}\,{=}\,1.5$.
SMM\,02399 is better modelled by a lower dust temperature
($T_{\rm d}\,{\simeq}\,40\,$K)
while SMM\,14011 is better modelled by a higher dust
temperature ($T_{\rm d}\,{\simeq}\,60\,$K).  This implies that the technique
of predicting the dust emission from the UV parameters is reasonably
sound if the dust temperature distribution is known.  However, the
850\mum\ flux density for W-MMD11 is under-predicted, unless the dust
temperature used is much lower (say $T_{\rm d}\,{=}\,32\,$K).  This is in
strong contrast to the results for the rest of the LBGs which would indicate
quite a hot $T_{\rm d}$ if we fixed all the conversion factors at typical
values.  Of course, different temperatures are not the only explanations
for why we detect W-MMD11 alone.
Another possibility is that the spatial distribution of the dust could
have given rise to a detection in emission, but less effect on the
absorption properties.

Although our targetted sample are not particularly bright in the
sub-mm, recent simulations (Adelberger \& Steidel~2000)
suggest that {\it some\/} of the LBGs in the total sample
must have large enough SFRs to be detected with SCUBA
(${>}\,400\,{\rm M}_\odot\,{\rm yr}^{-1}$).
The most likely galaxies to be detected might be those with
brighter apparent magnitudes and smaller
implied dust corrections, since errors in $R_{\rm s}$ are smaller
than errors in $G\,{-}\,R_{\rm s}$, and the corrected star-formation rates
are less sensitive to errors in $R_{\rm s}$ than in $G\,{-}\,R_{\rm s}$.
Although some galaxies in our sample are already of this type (our
detection W-MMD11 being one of them),
this possibility should be further checked in future with separate samples.
%It is still reasonable to assume that many of the LBGs in our SCUBA sample
%should be representative
%of the targets most likely to be detectable with SCUBA.
Also, as deep $K$-band photometry becomes available for the LBG samples, 
the $R\,{-}\,K$ colour may prove to be a useful selection criterion for 
sub-mm follow-up. Indeed, several SCUBA-selected galaxies, beyond those with
known redshifts discussed in section 4.3, are now believed
to have optical counterparts with very red $R\,{-}\,K$ colours
(Smail et al.~1999a) classifying them as Extremely Red Objects
(see for example Dey et al.~1998).

\section{Discussion}
There is currently a great deal of debate on reconciling
the various techniques for estimating the SFR using different wavelength
regimes, such as sub-mm, UV continuum or H$\,\alpha$ (see
for example: Kennicutt 1998; Ouchi et al.~1999; Cowie, Songaila \&
Barger~1999).
%Although we have tried to outline the uncertainties in the chain of
%estimation methods which we have used,
%the relative normalization and quantitative interpretation of
%the SFR remain essentially unknown. This is apparent in the results of
%our SFR calculations as presented in section 4.2.
Regardless of the precise relationships between SFR estimators, 
we can certainly say that the LBGs with the largest expected 
SFRs are apparently not excessively bright sub-mm emitters, and are
less luminous than the typical sub-mm sources discovered in a SCUBA `blank
field' survey \cite{SIB,Barger,Hughes,Lilly99}.

As discussed in section 4.2, if we assume that the SCUBA flux density
is proportional to
the UV-predicted flux density, then our data can be used to infer that the
UV-predicted flux densities are on average too high by factors of a few,
depending on the dust temperature.  For $T_{\rm d}\,{\la}\,50\,$K this
factor would be even higher, while for
For $T_{\rm d}\,{\ga}\,70\,$K, the UV predictions
are consistent with the measured SCUBA flux densities, within uncertainties.
By contrast, the detection (W-MMD11) in our sample is at a level
{\it above\/} that implied by the UV calculations, even for unusually
cool dust temperatures.

Since we currently have only a small sample of sufficiently deep integrations,
including one probable detection, we cannot conclude
from our results that {\it all\/} LBGs are likely to be weak sub-mm emitters.  
As discussed in section 3.3, recent results (Adelberger \& Steidel~2000)
suggest that the LBGs with the largest SFR correction factors
(e.g.~W-MMD109 in Table 1) are probably subject to larger UV photometric
and Malmquist-type errors, and have lower corrected SFRs when the
uncertainties are taken into account.  This could lead to a situation where
only a few of the galaxies in this sample truly have large SFRs, which would
also be consistent with our observations.

On the other hand, we should emphasize that there is no reason
a priori to assume that the high 
redshift galaxies have a similar relationship between far-IR/far-UV flux and
UV continuum slope to the local starburst population, which itself has large
scatter (Meurer et al.~1999). This may well play a role in
any discrepancy between our sub-mm data and predictions based on the UV.
A reasonable conclusion is therefore that the simplest predictions for
SCUBA flux densities for our targets are on the average overestimated by a
factor of order a few.  The precise reason for this is difficult, at
present, to ascertain.

\subsection{Relationship with sub-mm selected sources}

The central question remains: is there any overlap between samples selected
as LBGs and SCUBA-detected blank field sources, when they are at the same
redshift? 
The difficulty in obtaining accurate optical counterparts
and redshifts for the sub-mm sources means that the volume and
luminosity function of sub-mm
`blank field' surveys is still essentially unknown. Initial spectroscopic
follow-up (Barger et al.~1999; Lilly et al.~1999) has suggested that
a significant fraction of the SCUBA population does not lie at $z\,{>}\,2$.
However, use of the Carilli and Yun (1999)
relation of radio/far-IR flux to predict the redshifts for
a large sample of SCUBA sources (Smail et al.~1999b), has suggested that
the population may have a median value lying between
$z\,{=}\,2.5$--3. It is worth noting that the
detection of W-MMD11 in our sample currently represents the highest redshift
source detected with SCUBA, which is not an AGN.

This one detection suggests that with reasonable SCUBA integrations we might
expect to detect just those few LBGs that are far-IR brightest.
This would be consistent with `blank field' sources lying over
a wider range of
redshifts than that probed by the $U$-band drop-out selection technique,
where a few extreme specimens at any given epoch radiate strongly
in the sub-mm.
The fact that our high SFR Lyman break sample is on average
undetected in the sub-mm implies that probably only the very highest
star formers would constitute part of the blank-field sub-mm sources.
LBGs might also be harder to detect if they had higher dust temperatures
(as depicted in Fig.~4).

We now consider scenarios which are consistent with our data
and in which the population of LBGs is indeed related
to the population of sub-mm selected  sources.  If none of these
proves plausible one could always consider the possibility that the
relationship between far-IR and far-UV flux, and UV continuum slope to
be different at high redshift than it is for local starburst galaxies.

If many of the blank field sub-mm sources are in fact star-forming,
merging galaxies at even higher redshifts ($z\,{\sim}\,4$--5), as yet
undiscovered in optical surveys, they might represent an era when
the star formation was much more vigorous and short lived.
Recent results (Steidel et al.~1998) have revealed more $z\,{\sim}\,4$
LBGs than implied by number counts and modelling of data from the HDF.
These are not expected to be strong sub-mm emitters, since a similar
colour range to the $z\,{\sim}\,3$ population is observed. Massive merging
fragments are now thought to be responsible for the prodigious SFRs observed
in many SCUBA sources (Blain et al.~1999a), perhaps implying more
SCUBA-bright sources during the period of most merging activity.
But certainly, accurate prediction
of sub-mm properties from optical properties will await a more complete
understanding of the galaxy formation process.

Recent observations have revealed possible analogues to the `blank field'
discovered sub-mm sources \cite{Iviplus7,ChaSLBF},
where the sub-mm flux density is likely to be dominated by AGN heated dust.
It is possible that
spectroscopic follow up of `blank field' sources may reveal that more are
AGN dominated, implying a large population of high redshift dusty quasars.
Indeed consideration of x-ray results suggest that a significant fraction
of the submm sources could involve active galaxies \cite{AlmLB}.
If the primary engines powering these sub-mm sources are AGNs, then there
may indeed be little relation between these galaxies and the Lyman break
population.

Both the LBGs and SCUBA-selected sources are thought to be associated
with elliptical galaxies in the process of formation. They can be
reconciled if the dust content of young galaxies is coupled to their
mass or luminosity, with more massive galaxies being dustier
\cite{dickinson98}.  The most massive young elliptical galaxies could
then be associated with the sub-mm sources, while the Lyman break
population would be identified with less massive ellipticals and
bulges. To verify this possibility, accurate identifications and
dynamical mass estimates of the sub-mm galaxies are required, which
may have to wait for the next generation of sub-mm interferometers to
detect CO lines (see e.g.~Stark et al.~1998).

\subsection{Contribution to the far-IR background}

One assessment of the significance of the sub-mm emission of
LBGs is to estimate their contribution to the IR
background at 850\mum.  We can calculate this using the average implied SFRs
obtained using the same methods outlined here, together with the surface
density estimate for $z\,{\sim}\,3$ LBGs. We find that typical
redshift 3 LBGs then account for about 0.2 per cent of the 850\mum\
background estimates \cite{Fixsen,Lagache}.
{}From our SCUBA observations, we find no evidence that the LBG population
contributes more than this.  However, W-MMD11 is actually much
{\it more\/} SCUBA-bright than predicted, and so the full
contribution will depend on how common such objects are in the
LBG population, which we certainly cannot estimate from our small sample.

A significant problem with this estimate is that there is a strong bias,
in present samples, against selecting just the sorts of LBG candidates
which might dominate the far-IR background.
This bias arises from several causes: highly redenned objects have colours
satisfying our $UGR$ selection criteria over a much shorter
range of redshifts than for bluer objects; at fixed SFR 
a much smaller fraction of reddened than unreddened objects
will satisfy the $R_{\rm s}\,{<}\,25.5$ magnitude cut; and red objects tend
to be so faint in $G$ that they are difficult to detect and recognize
as LBGs down to the $R_{\rm s}$ limit.  We certainly know that highly redenned
LBGs are very underrepresented in the current sample, and it
is impossible to accurately estimate the
850\mum\ flux for the whole LBG population without fully understanding the
relevant selection biases.  Because of these biases, it is still possible
that the $z\,{\sim}\,3$ LBG population contributes a significantly higher
portion of the background radiation at these wavelengths.

\section{Conclusions}

\begin{enumerate} 
\item On average we find that the 850\mum\ flux density
must be at least 2 times lower than the simplest predictions
obtained from the UV colours.
This could be accounted for by a combination of photometric errors,
uncertainties in $T_{\rm d}$, $\beta_{\rm d}$,
or the estimates of $L_{\rm bol}$ from the
rest-frame UV and far-IR wavelengths or
from the scatter in the UV-slope/far-IR relation.
Our sample also had some bias against the most highly reddened objects.
\item In the case of the detection of W-MMD11, the flux density is a factor
${\sim}\,5$ times greater than predicted by the UV for dust
temperatures $T_{\rm d}\,{>}\,50\,$K.
\item The similarities in the properties of W-MMD11 and SCUBA-selected
sources of known high redshift, SMM J02399$-$0136 and SMM J14011+0252,
suggest that the large UV-implied SFR in conjunction with a red $R\,{-}\,K$
colour may be a good indicator of significant sub-mm flux density.
The comparison may indicate that the
prediction of far-IR flux density from UV colours is fairly reliable, even
for quite reddened objects, provided that parameters, such as the dust
temperature are known reasonably well.
\item Estimates for the contribution of the $z\,{\sim}\,3$ LBGs to the far-IR
background give around 0.2 per cent.  Our non-detections certainly
provide no evidence that the contribution is significantly higher than this.
However, given the fact that our one detection has
considerably lrager sub-mm flux than predicted, and that there are selection
biases against highly redenned objects, it is difficult to estimate
precisely the overall contribution to the far-IR background.

\end{enumerate} 

Our detection of W-MMD11 certainly indicates that there is some overlap
between $z\,{\sim}\,3$ LBG and SCUBA galaxies.  However, our small sample makes
it hard to draw any firmer conclusions about how great the overlap might be.
It also remains to be seen whether the galaxies contributing to the bulk of
the far-IR background, over the full range of redshifts, would be also be
selected by the UV-dropout technique.
Further progress will require significantly more telescope time to improve
the sub-mm limits and detections.  We have shown that targetted SCUBA
photometry is a useful approach here.
With larger sub-mm data sets, and selection of
LBG samples in different ways, it should be possible to
test the predictive power of the far-UV for luminous
star forming galaxies at high redshift, and to more fully investigate
the role of dust in the galaxy formation process.

%%%%%%%%%%%%%%%%%%%%%%%%%%%%%%%%%%%%%%%%%%%%%%%%%%%%%%%%%%%%%%%%%%%%%
\section*{ACKNOWLEDGMENTS}

This work was supported by the Natural Sciences and Engineering
Research Council of Canada.
The James Clerk Maxwell Telescope is operated by
The Joint Astronomy Centre on behalf of the Particle Physics and
Astronomy Research Council of the United Kingdom, the Netherlands
Organisation for Scientific Research, and the National Research
Council of Canada.
We would like to thank the staff at JCMT for facilitating
these observations.  We are also grateful to Remo Tilanus and his
colleagues for their willingness to discuss their own LBG observations.
%%%%%%%%%%%%%%%%%%%%%%%%%%%%%%%%%%%%%%%%%%%%%%%%%%%%%%%%%%%%%%%%%%%%%

%[MANY REFS ARE UNCITED IN THE TEXT SO FAR]

\end{document}